\documentclass[12pt]{iopart}

\usepackage{amssymb}
\usepackage{wasysym}
\usepackage{graphicx}
\usepackage{xcolor}
\usepackage{siunitx}
\newcommand{\red}[1]{\textcolor{black}{#1}}

\usepackage[displaymath]{lineno}
\usepackage[pdftitle={Yongxiang Huang: JHydrolb},bookmarks=true,citecolor=red,colorlinks=false]{hyperref}

\begin{document}

\title[SST Cascade  in the Oceanic System]{Cascade and Intermittency of the Sea Surface Temperature in the Oceanic System}

\author{Yongxiang Huang$^{1}$ \& Lipo Wang$^{2}$}

\address{$^{1}$State Key Laboratory of Marine Environmental Science \& College of Ocean and Earth Sciences, Xiamen University, Xiamen 361102, PR China\\
$^{2}$UM-SJTU Joint Institute, Shanghai JiaoTong University, Shanghai 200240, RP China}
\ead{\mailto{yongxianghuang@gmail.com},\mailto{lipo.wang@sjtu.edu.cn}}
\vspace{10pt}
\begin{indented}
\item[]July 2018
\end{indented}

\begin{abstract}
In this paper, we analyze the sea surface temperature obtained from the global drifter program. The experimental Fourier power spectrum shows a two-decade power-law behavior as $E_{\theta}(f)\propto f^{-7/3}$ in the frequency domain. Dimensional argument suggests a  two-dimensional-like Lagrangian forward cascade, in which the enstrophy dissipation $\epsilon_{\Omega}$ is involved. Using the Hilbert-Huang transform and multi-level segment analysis, the measured high-order statistics and the corresponding singularity spectrum confirm the existence of the intermittency with a measured intermittency parameter $\mu_{\theta}\simeq0.10$, which is much weaker than the prediction by the conventional structure function method.
 \end{abstract}

\noindent{\it Two-dimensional turbulence, Sea surface temperature, Lagrangian statistics, Intermittency\/}

\submitto{\PS}
\maketitle
\section{Introduction}
As a typical complex dynamical system, turbulence with diverse configurations shows a significant scale invariant property  \cite{Frisch1995,Thorpe2005book}.  \red{Generally, scale invariance implies features or universal laws which remain invariant with respect to the scale. For instance,  the celebrated  Kolmogorov 1941 theory (hereafter referred as K41) has been} put forward to understand quantitatively the small-scale fluctuation of the Eulerian velocity in the framework of Richardson-Kolmogorov energy cascade \cite{Sreenivasan1997,Frisch1995}. The scale invariance is characterized by the Kolmogorov $5/3$-law as $E(k)\sim k^{-5/3}$, when $k$ lies in the so-called inertial range $k_L\ll k\ll k_{\eta}$, where $k_{\eta}$ is the Kolmogorov scale and $k_L$ is the forcing scale; or equivalently the high-order structure-function, $S_q(\ell)=\langle \Delta u_{\ell}(x)^q\rangle \sim \ell ^{\zeta(q)}$, when $\ell$ lies in the inertial range $\eta\ll\ell\ll L$, where $\zeta(q)=q/3$. 
The anomaly scaling, i.e. the high-order scaling exponent $\zeta(q)$ deviates from the non-intermittent K41 predictions ($\zeta(q)=q/3$), is discovered experimentally in turbulent shear flows by Anselmet et al., \cite{Anselmet1984}. \red{Intermittency originates as a burst of energy dissipation field, where a huge fluctuation of dissipation event is observed. It is further considered as a result of the nonlinear interaction in the Navier-Stokes equation \cite{Frisch1995}.}
Inspired by their experiment observation, Parisi \& Frisch \cite{Parisi1985} introduced the multifractal concept to explain the anomaly scaling \cite{Benzi1984}. Identified in different types of turbulent flows, multifractality is recognized as a common  feature of complex dynamical systems, in which a bunch of freedoms interact with each other, resulting in scale invariance over the inertial range.

In the real geophysical system, because of the very large characteristic scales and thus the Reynolds number, turbulence phenomena are relevant, in which many spatial and temporal scales coexist and interact with each other. Meanwhile, geophysical flows, either oceanic or atmospheric, have their special and complex features \cite{Lovejoy2013Book}. For example, the marine turbulence is driven by the solar radiation either directly or indirectly with a typical daily and annual cycle \cite{Deser2010ARMS}. A continuous scale range then presents at least in between the daily and annual cycles, which has been confirmed in many observed geophysical data, such as the atmospheric temperature \cite{Finnigan2000ARFM,Shaw2003ARFM} and ocean currents \cite{Wunsch2004ARFM}. Therefore, the turbulence statistics and the multiscaling features are crucial in understanding the ocean properties, for instance, the velocity, temperature and biomass concentration, etc. Specifically, because of the configuration confinement, geophysical turbulence assumes both two-dimensional (2D) and three-dimensional (3D) properties at different (spatial and temporal) scales~\cite{Lovejoy2009}.

The sea surface temperature (SST) is relevant not only in the ocean dynamic system, but also in the climate process.
This is because more than $70\%$ of the earth surface is covered by the ocean, and at the same time the heat
capability of sea water is much larger than of the air \cite{Deser2010ARMS}. Hence the SST dominates the heat
transport during the atmosphere-ocean interaction \cite{Hendon2005air,Gill1982atmosphere}. Most of the previously
studies of SST focus on the climate aspect, which is mainly related to the global warming by checking the trend of the
annual averaged global mean SST \cite{Deser2010ARMS,Chen2014Science}. However, the dynamic perspective of
SST, especially multiscaling and multifractality, is seldom studied. Nieves et al., \cite{Nieves2007GRL} applied a microcanonical multifractal formalism \cite{Turiel2005PRL} to both SST and chlorophyll concentration obtained from satellite images derived from Aqua-MODIS ocean color sensor. Due to the advection of the quasi-2D oceanic turbulence both SST and chlorophyll concentration show the same singularity spectra \cite{Isern2007JGR}. Abraham \& Bowen \cite{Abraham2002Chaos} observed a scaling exponent around $\beta=2.44$ from the Fourier power spectrum of SST. Differently from the 2D structure of the SST obtained from satellite remote sensing, Renosh, Schmitt \& Loisel \cite{Renosh2015PLOS} reported a scaling exponent $\beta=1.8$ for the Fourier power spectrum of SST. Carbone, Gencarelli \& Hedgecock  \cite{Carbone2016PRE} studied the scaling behavior of SST provided by the Lagrangian drifter in the Agulhas return current. A Kolmogorov-Landau type spectrum $E(f)\propto f^{-2}$ was observed in the frequency range $2\times 10^{-5} \sim 5\times 10^{-4}$ Hz, corresponding to a time scale $0.6\sim 14\,$hours. Lin, Zhuang \& Huang reported a dual-power behavior in the Gulf of Mexico in time domain with scaling exponent $\beta=1.59$ for the time scale larger than 1 day, and $\beta=2.89$ for the scale smaller than 1 day \cite{Lin2017SR}.  
 Such discrepancy implies the extremely complex physics of ocean turbulence.

There are a number of methodologies to quantify the multiscaling or multifractal property of a dynamical process.
For example, the classical structure-function (SF) \cite{Frisch1995}, wavelet-based methods (e.g., wavelet leaders, wavelet transform modulus maxima)
\cite{Muzy1993PRE,Lashermes2008EPJB,Huang2011PRE}, detrended fluctuation analysis \cite{Kantelhardt2002a}, Hilbert-based
method~\cite{Huang2008EPL,Huang2013PRE} and multi-level segment analysis~\cite{Wang2015JSTAT}.
It is important to consider the applicability of different methods. The existing work has demonstrated that SF is strongly influenced by energetic structures \cite{Davidson2005PRL,Schmitt2016Book}, such as ramp-cliff structure in the passive scalar field~\cite{Huang2011PRE}, and vortex trapping event in the Lagrangian turbulence~\cite{Huang2013PRE,Falkovich2012PoF}. The reason is that SF mixes the large- (known as infrared effect) and small-scale (known as ultraviolet effect) information \cite{Davidson2005PRL,Huang2011PRE}. The detrended fluctuation analysis suffers from the same problem~\cite{Huang2011PRE}. The wavelet-based method can be influenced by the nonlinear
property of the data, namely high-order harmonic problem leading the extracted multifractal spectrum biased~\cite{Huang2011PRE}.

Therefore, to characterize the appropriate multifractal and scaling properties of SST is important to understand the complex ocean turbulence system. In this paper, the Hilbert-Huang transform (HHT) and multi-level segment analysis (MSA)~\cite{Wang2015JSTAT} are introduced for the analysis and results will then be compared and explored in details.

\section{Data Presentation and Methodologies}
\subsection{SST from the Global Drifter Program}

The SST data used in this study is obtained from the Global Drifter Program (GDP), which is the principle component of the Global Surface Drifting Buoy Array, a branch of National Oceanic and Atmospheric Administration's  Global Ocean Observing System. GDP provides an accurate and globally dense set of in-situ observations of sea flow parameters. The temperature is measured with instantaneous sampling at every 6-hour from
each drifter.
Denote the temperature obtained from the $i$th drifter at location $(x,y)$ and time $t$ as $\theta_i(x,y,t)$. The measurement accuracy of the thermistor composite (YSI type 44018 or equivalent) is within $0.1\,^{\circ}\mathrm{C}$ (e.g., saying $0.05\,^{\circ}\mathrm{C}$)~\cite{Castro2012JGR}. The overall averaged temperature from all the drifters is
$\tilde{\theta}=\langle\theta_i(x,y,t)
\rangle_{i,x,y,t}=20.1\,^{\circ}\mathrm{C}$, with a standard deviation $\theta_{\mathrm{r.m.s}}=8.1\,^{\circ}\mathrm{C}$.
We have also calculated the time averaged temperature $\tilde{\theta}(X,Y)=\langle \theta_i(x,y,t)\vert_{x=X,y=Y}\rangle_{i,t}$, which is consistent with other observations \cite{Castro2012JGR}.
Due to many reasons, drifters have their limited life time span (i.e. the persistent life $T$). Thus the data length from each drifter varies
with missing data at some spots. The mean drifter persistent life $T$ and its standard deviation are  respectively $\langle T\rangle\simeq 370\,$ and $\sigma\simeq 360\,$ days. To ensure the measured data to be representative, new drifters need to be added to keep the total number above some certain
level. Here we consider the measurement span from 1 Jan. 2000 to 1 Jan. 2012.

\begin{figure}
\centering
\includegraphics[width=0.85\linewidth,clip]{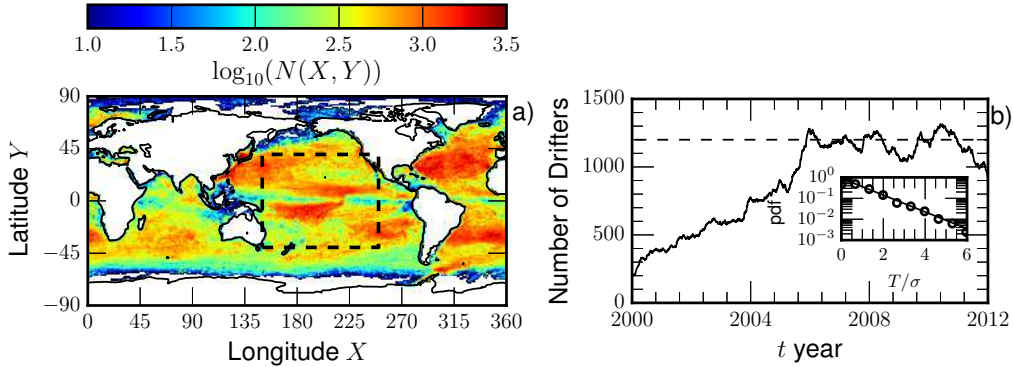}
\caption{a) Global spatial  distribution of Lagrangian drifters \red{with a spatial resolution of one degree both latitudinally and longitudinally}. The concentration of the Lagrangian drifter is partially due to the flow of the ocean current. For display convenience, the number $N$ has been taken its logarithm. A square (dashed line) indicates a area $x\in[150,250]$ and $y\in [-40,40]$ to exclude the continent boundary influence of calculating fractal dimension, see Fig.\,\ref{fig:FD}. b) The time evolution of the number of Lagrangian drifters on the time span 1st Jan. $2000$ to 1st Jan. $2012$.  The number $N$ approximates to a constant 1,200 since 1st Jan. $2006$. The inset shows the pdf of the life of Lagrangian drifters, in which the $\sigma\simeq 360$ is the standard deviation. An exponential law is observed with  a scaling exponent $0.40\pm0.02$ in the range $0.5\le T/\sigma\le 5$. }
\label{fig:drifter}
\end{figure}

Figure \ref{fig:drifter}\,a) shows the number distribution $N(X,Y)$ of these drifters. For display convenience, $N(X,Y)$ has been taken its logarithm. Visually, several patches are observed, indicating a clustering of drifters. \red{This is partially due to the flow topology of ocean current, and partially the initial release location of the drifters, which is associated with the region of interest.} The evolution of the drifter number is shown in Figure \ref{fig:drifter}\,b). The drifter number increases almost linearly from around 200 to 1200 from 1 Jan. 2000 to 1 Jan. 2006, and then keeps almost constant $\sim1200$ (the dashed line) after 1 Jan. 2006. The inset of Figure \ref{fig:drifter}\,b) shows the probability density function (pdf) of the drifter persistent life time $T$. An exponential law is observed in the range $0.5\le T/\sigma \le 5$ with a scaling exponent $\red{-0.40\pm0.02}$. \red{Here the uncertainty (resp. error bar) is provided by the $95\%$ fitting confidence level to feature the power-law behavior. }

\begin{figure}
\centering
\includegraphics[width=0.65\linewidth,clip]{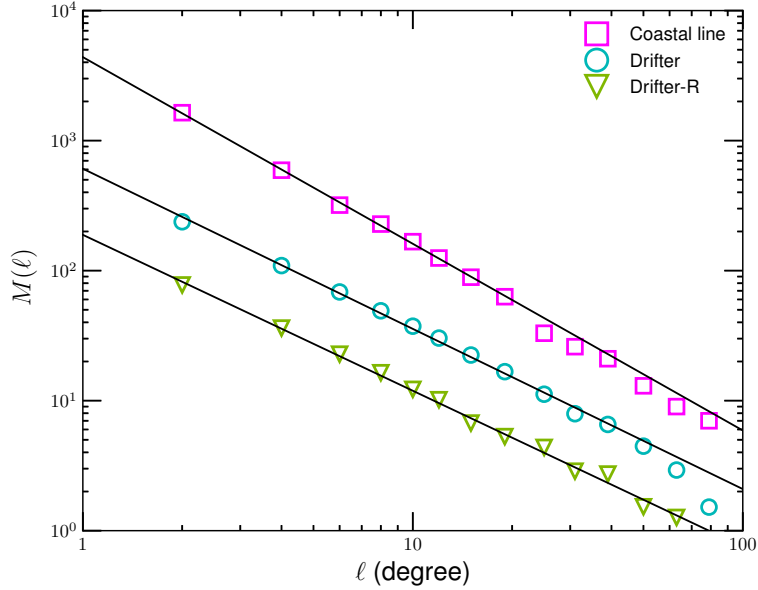}
\caption{Fractal dimension of the spatial distribution of Lagrangian drifters ($\ocircle$) with $D=1.18\pm0.03$. For comparison, the fractal dimension of the coastal line is shown as $\square$ with $D_c=1.44\pm0.04$. The value for the area $x\in $[150,250] and $y \in$[-30,30] is also calculated as $D_R=1.20\pm0.07$.  
}
\label{fig:FD}
\end{figure}

To characterize more precisely the spatial distribution of these drifters, the fractal dimension is calculated via the box-counting method, i.e.,
\begin{equation}
M(\ell)\sim \ell^{-D}, \label{eq:fractal}
\end{equation}
in which $M(\ell)$ stands for the number of the counting boxes with drifters inside, $D$ is the fractal dimension. The box size $\ell$ is simplified by $\ell \sim \sqrt{d\varphi\times d\phi}$, where $d\varphi$ and $d\phi$, the change of longitude $\varphi$ and latitude $\phi$, are both set as one degree. Because of the spherical rather than planar structure of the earth surface, boxes need be demarcated by the spherical coordinates via the Mercator Projection.
We calculate the counting boxes for each day and take the average over all the time span. Figure \ref{fig:FD} shows the measured $M(\ell)$ for both the coastal line ($\square$) and the Lagrangian drifter ($\ocircle$). Power-law behavior is observed in the large scale range. The fitted fractal dimension is respectively $D=1.18\pm0.03$ for drifters and $D_c=1.44\pm0.04$ for the costal line. The $D_c$ value is consistent with the reported result in other literatures \cite{Mandelbrot1975PNAS}, indicating the reliability of the present measurement. To justify the simplification of $\ell$ and understand the potential bias from the Mercator Projection and the continent boundary, a sub-region close to the equator, as shown in Fig. \ref{fig:drifter}\,a) by the dashed line ($x\in[150,250]$ and $y\in [-40,40]$), is also considered. The corresponding $D$ is found to be $1.20\pm 0.07$ with negligible difference.

\begin{figure}
\centering
\includegraphics[width=1\linewidth,clip]{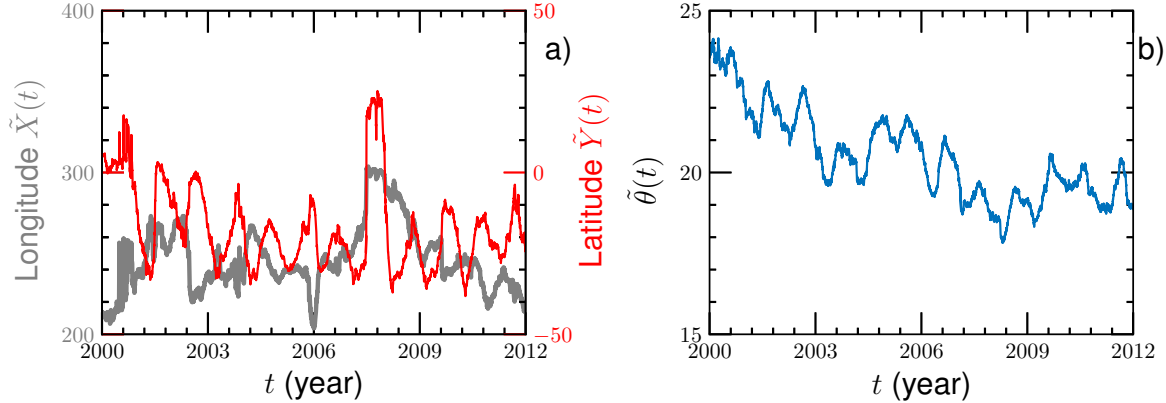}
\caption{ a) Temporal variation of the mass center defined by equation \ref{eq:masscenter}. Visually, an annual cycle is observed for
$\tilde{Y}(t)\rangle$. Such fluctuation indicates that even after spatial average SST still preserves to be Lagrangian.
b) The drifter averaged temperature $\overline{\theta}(t)=\langle \theta_i(x,y,t)\rangle_{i,x,y}$ as a function of time. An annual  cycle can clearly be observed. 
}
\label{fig:globalaverage}
\end{figure}

The fact that $D$ is close to 1 indicates that the averaged quantities with respect to measurements from all the drifters, such as the averaged temperature introduced below, can not be treated as a global average.  \red{More precisely, it might be  considered as a line measurement of SST. This can be further characterized by  the mass center of these drifters}, i.e.,
\begin{equation}
R(\tilde{X}(t),\tilde{Y}(t))=\min_{X,Y}\left\{  R \right\},\,R(X,Y,t)=\langle (X_i(t)-X,Y_i(t)-Y)\vert_i \rangle\label{eq:masscenter}
\end{equation}
\red{where $\langle\,\cdot\, \rangle$ means average, and $(X_i-X,Y_i-Y)\vert_i$ is the great circle distance between geo-position $(X_i,Y_i)$ and $(X,Y)$.   With a uniform distribution of drifter without complex boundary, we have a flat $R$, we then define  $\tilde{X}(t)= 180$ and $\tilde{Y}(t)= 0$ with the spherical coordinates.}
Figure \ref{fig:globalaverage}\,a) shows large variation, implying the movement of the overall drifters still (partially) preserves the Lagrangian property since $D$ is close to 1. Physically this interesting feature is relevant to the large scale tracing of the drifters. The obtained temperature series is more like a kind of large-scale filtering, i.e. neither Lagrangian nor Eulerian. More details can be referred to Ref.~\cite{Castro2012JGR}.

The temperature averaged from GDP collection is defined as
\begin{equation}
\tilde{\theta}(t)=\langle \theta_i(x,y,t\rq{})\vert t\rq{}=t \rangle _{i,x,y}.
\label{eq:ASST}
\end{equation}
In the same vein that the fractal dimension of the drifters $D=1.18$, much smaller than 2, drifters follow the large-scale movement in ocean turbulence without inertial effect, and can preserve the large time (spatial) coherent structure. We argue here that the average operator ${\langle{\cdot}\rangle}$ in equation\,\ref{eq:ASST} behaves as a lower-pass filter \cite{Lumpkin2010JGR}. Thus $\theta_i(x,y,t)$ is the Lagrangian or partially filtered Lagrangian temperature~\cite{Thorpe2005book}, whatever it is inclined to be active or passive, 2D or 3D dominated \cite{Warhaft2000,Boffetta2012ARFM,Bouchet2012PhysRep}.

The variation of $\tilde{\theta}(t)$ with respect to $t$ is shown in Figure \ref{fig:globalaverage}\,b) \red{totally with 17,532 data points}. A clear annual cycle appears because of the external influences, e.g. earth rotation and earth revolution. Overall from 2000 to 2006 $\tilde{\theta}(t)$ first decreases and then stays around $19\,^{\circ}\mathrm{C}$ after 2006. 
 The present study is an attempt to understand to global flow turbulence features based on the real experimental from GDP. Especially we focus on the multi-scale and multifractality of the \lq{}filtered\rq{} temperature $\tilde{\theta}(t)$ to address some important features with common interests.

\subsection{Methodologies}

\subsubsection{Structure Function}
The structure-function method was first proposed by Kolmogorov in K41 and widely used since then to characterize the scale similarity \cite{Frisch1995}. For the $\tilde{\theta}$ case the $q$th-order SF is defined as (in the temporal domain for Lagrangian statistics)
\begin{equation}
S_q(\tau)=\left\langle \left(  \Delta \tilde{\theta}_{\tau}(t)\right)^q \right\rangle\sim \tau^{\zeta^S_{\tilde{\theta}}(q)},\label{eq:qSF}
\end{equation}
in which $\Delta \tilde{\theta}_{\tau}(t)=\vert \tilde{\theta}(t+\tau)-\overline{\theta}(t)\vert$
is the temperature increment, $\tau$ is the separation scale and
$\zeta_{\tilde{\theta}}^S(q)$ is the SF scaling exponent.

It has been reported by several authors that SF analysis may mix the information between large- and small-scale structures, also known as infrared and ultraviolet effects \cite{Davidson2005PRL,Wang2006JFM,Huang2013PRE,Tan2014PoF}. Such kind of mixing becomes more serious when the
energetic structure exists or $\beta\ge2$, where $\beta$ is the slope of the Fourier power spectrum, e.g., $E(f)\sim f^{-\beta}$ ~\cite{Huang2013PRE,Tan2014PoF}. Typical examples include the passive scalar turbulence with a ramp-cliff structure, the active scalar turbulence in Rayleigh-B\'{e}nard convection with a large-scale circulation, the vortex trapping event in the Lagrangian turbulence \cite{Huang2013PRE}, the forward enstrophy cascade in 2D turbulence \cite{Tan2014PoF}, and the observation data from geosciences with annual cycle as to be shown in this paper \cite{Huang2009Hydrol}.

%

\begin{figure}
\centering
\includegraphics[width=1\linewidth,clip]{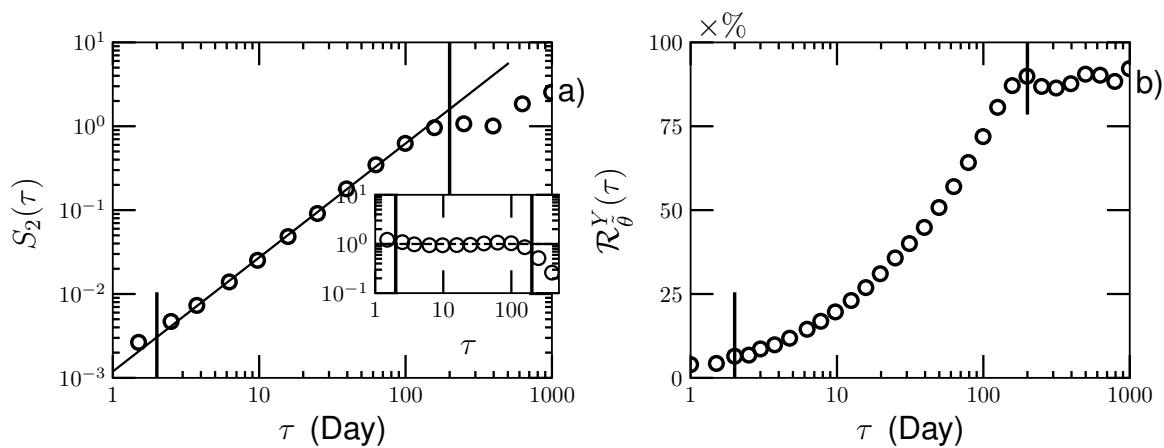}
 \caption{a) The experimental second-order structure-function $S_2(\tau)$ in the range $2<\tau<100\,$Dtay (corresponding to $0.01<f<0.5\,$Day$^{-1}$) with a scaling exponent $\zeta_{\tilde{\theta}}^S(2)=1.36\pm0.02$, which is slightly larger than the prediction by the Fourier spectrum, i.e. $2.32-1=1.32$. \red{To clarify the scaling behavior, the first few points are ignored here}. The inset shows the compensated curve to emphasize the observed power-law behavior. b) The relative cumulative function $\mathcal{R}_{\theta}({f_M},\tau)$ with $f_M=f_Y=1/T_Y$. Note that the second-order SF is strongly influenced by the scales larger than annual cycle. The power-law range predicted by the Fourier analysis is indicated by a vertical line. }
 \label{fig:2SF}
 \end{figure}

The corresponding second-order SF $S_2(\tau)$ is shown in figure\,\ref{fig:2SF}\,a), with $\zeta_{\tilde{\theta}}^S(2)=1.36\pm0.02$ in the time scale range $2<\tau<100\,$Day.  
The measured $\zeta_{\tilde{\theta}}^S(2)$ is slightly larger than the value deduced by the Fourier spectrum, i.e., $\zeta_{\tilde{\theta}}^S(2)=\beta_{\tilde{\theta}}-1=2.32-1=1.32$.
To understand the influence of the annual cycle, we provide here a Fourier-based scale analysis as in Refs. \cite{Huang2011PRE,Huang2013PRE,Tan2014PoF}.
The second-order SF $S_2(\tau)$ can be related with the corresponding Fourier power spectrum $E_{\tilde{\theta}}(f)$ via the Wiener-Khinchin theorem, i.e.,
\begin{equation}
S_2(\tau)=\int_0^{+\infty} E_{\tilde{\theta}}(f)(1-\cos(2\pi f \tau)) df,
\label{eq:WKT}
\end{equation}
in which the $E_{\tilde{\theta}}(f)$ is the experimental Fourier power spectrum of $\tilde{\theta}$, and $\tau$ is the time separation scale.
For a scaling process, e.g., $E_{\tilde{\theta}}(f)\sim f^{-\beta_{\tilde{\theta}}}$ with $1<\beta_{\tilde{\theta}}<3$ \cite{Huang2011PRE}, the second-order SF has a scaling as $S_{2}(\tau)\sim \tau^{\zeta_{\tilde{\theta}}(2)}$, where $\zeta_{\tilde{\theta}}(2)=\beta_{\tilde{\theta}}-1$.
However, the Wiener-Khinchin theorem implies that except for the case $f=n/\tau$, $n=0,1,2\cdots$, all Fourier components have
contribution to $S_2(\tau)$. With the increase of $\beta_{\tilde{\theta}}$, SF becomes more influenced by the low-frequency (i.e. large-scale) part, which can be quantified by the following relative cumulative function $\mathcal{R}_{\tilde{\theta}}({f_M},\tau)$ measuring a relative contribution from frequency band $[0, f_M]$:
\begin{equation}
\mathcal{R}_{\tilde{\theta}}({f_M},\tau)=\frac{\int_0^{f_M} E_{\tilde{\theta}}(f)(1-\cos(2\pi f \tau)) d f}{\int_0^{+\infty} E_{\tilde{\theta}}(f)
(1-\cos(2\pi f \tau)) d f}\times 100\%.
\label{eq:cumulative}
\end{equation}
The special case $f_M=1\,$Year$^{-1}$ provides a quantitatively characterization of the relative contribution from the the time scale $t\ge 1\,$Year since the strong annual cycle is observed. Numerically $\mathcal{R}_{\tilde{\theta}}$ increases from $5\%$ to $70\%$ in the range $2<\tau<100\,$Day, showing a strong influence of the large-scale variation, i.e. $f\le 1\,$Year$^{-1}$.
In other words, the SF scaling is nearly dominated by the energetic large-scale part. We argue here that not only
the second-order SF $S_2(\tau)$ but also  the high-order $S_q(\tau)$ cases, are strongly influenced by large-scale
motions.

\subsubsection{Hilbert-Huang Transform}
The general view of the HHT is that the signal from the real world consists different scales simultaneously \cite{Huang1998EMD,Rilling2003EMD}. Then the Intrinsic Mode Function (IMF) is proposed to represent a mono-scale, which satisfies the following conditions: (\romannumeral1) the difference between
the number of local extrema and the number of zero-crossings must be zero or
one; (\romannumeral2) the running mean value of the envelope
defined by the local maxima and the envelope defined by the local minima is
zero \cite{Huang1999EMD}. A sifting algorithm  is  designed to decompose  a given signal
 into several  IMF modes.
 For a given time series  $x(t)$, the
first step of the sifting process is to extract all the local  maxima (resp.
minima) points. The upper  envelope $e_{\max}(t)$ and the lower envelope
$e_{\min}(t)$ are then constructed, respectively, for the local  maxima and minima
points  by using  a cubic spline  algorithm or other algorithm \cite{Huang1998EMD,Rilling2003EMD}. The running mean between these two
envelopes is defined as
 \begin{equation}
 m_{1}(t)=\frac{e_{\max}(t)+e_{\min}(t)}{2}.
 \end{equation}
Thus the first component is  estimated by
\begin{equation}
h_1(t)=x(t)-m_{1}(t).
\end{equation}
Ideally,  $h_1(t)$ should be an IMF as expected.  In practice,  $h_1(t)$  may  not satisfy
 the above mentioned conditions. The function $h_1(t)$ is then
taken as a new time series, and  this sifting process is repeated $j$ times, until
$h_{1j}(t)$ is an IMF
\begin{equation}
h_{1j}(t)=h_{1(j-1)}(t)-m_{1j}(t).
\end{equation}
The first  IMF component $C_{1}(t)$ is then written as
\begin{equation}
C_{1}(t)=h_{1j}(t),
\end{equation}
and the residual $r_1(t)$ as
\begin{equation}
r_{1}(t)=x(t)-C_{1}(t).
\end{equation}
The sifting procedure is then repeated
on the residual,  until  $r_n(t)$ becomes a monotonic function or at most has one
local extreme point. This means   that no more  IMF can be extracted  from
$r_n(t)$.  There are finally  $n-1$
IMF modes with one residual $r_n(t)$. The original signal $x(t)$ is    rewritten
at the end of the process as
\begin{equation}
x(t)=\sum_{i=1}^{n-1}C_i(t)+r_{n}(t).
\end{equation}
To guarantee that the  IMF modes retain enough physical sense, a certain stopping criterion has to be introduced to stop the sifting process properly. Different types of stopping criteria have been introduced by several authors \cite{Huang1998EMD,Huang1999EMD,Rilling2003EMD}. The first stopping criterion is a  Cauchy-type convergence criterion. A standard
deviation (SD),  defined for two successive sifting processes is written as
\begin{equation}
\mathrm{SD}=\frac{\sum_{t=0}^{T}\vert
h_{i(j-1)}(t)-h_{j}(t)\vert^2}{\sum_{t=0}^{T} h_{i(j-1)}^2(t)}.
\end{equation}
If a calculated SD is smaller than a given value, then the sifting stops,
and gives an IMF.
A typical value   $0.2\sim 0.3$ has been proposed based on Huang et al.\rq{}s experiences \cite{Huang1998EMD,Huang1999EMD}.
Another widely used criterion is based on three thresholds $\alpha$, $\theta_1$, and
$\theta_2$, which are designed  to guarantee globally small fluctuations meanwhile taking into account locally large excursions \cite{Rilling2003EMD}. The mode amplitude and evaluation function are
\begin{equation}
a(t)=\frac{e_{\max}(t)-e_{\min}(t)}{2},
\end{equation}
and
\begin{equation}
\sigma(t)=\vert m(t)/a(t)\vert.
\end{equation}
Therefore the sifting is iterated until $\sigma(t)<\theta_1$ for some prescribed
fraction $1-\alpha$ of the total duration, while $\sigma(t)<\theta_2$ for
the remaining fraction. Typical values proposed by Rilling, Flandrin \& Gon{\c{c}}alv\`es  \cite{Rilling2003EMD}  are
$\alpha \approx 0.05$, $\theta_1
\approx 0.05$ and $\theta_2 \approx 10 \,\theta_1$, respectively based on their experience.
A maximal iteration number (e.g., $300$) is also chosen to avoid over-decomposing the time series.



In the second step, the Hilbert transform is applied to each IMF mode $C_i(t)$ to construct an analytical function, i.e.,
\begin{equation}
\tilde{C}_i(t)=P\frac{1}{\pi}\int_0^{+\infty}\frac{C_i(t\rq{})}{t-t\rq{}} d t\rq{},\quad C^{A}_i(t)=C_i(t)+j\tilde{C}_i(t),
\label{eq:HilbertT}
\end{equation}
in which $P$ stands for the Cauchy principle value \cite{Huang1999EMD}. The following  phase function, amplitude functionare respectively defined as, i.e.,
\begin{equation}
\mathcal{A}_i(t)=[C_i(t)^2+\tilde{C}_i(t)^2]^{1/2},\quad \phi_i(t)=\arctan\left(\frac{\tilde{C}_i{t}}{C_i(t)} \right)
\end{equation}
 The corresponding  instantaneous frequency is then written as, i.e.,
\begin{equation}
 \omega_i(t)=\frac{1}{2\pi}\frac{d \phi_i(t)}{d t},
 \end{equation}
Note that the Hilbert transform is a singularity transform and the first-order derivative of the phase function is used to define the instantaneous frequency $\omega_i(t)$. Therefore the HHT method is very capable to describe the local features in both physical and spectral domains.
With the extracted instantaneous frequency $\omega_i(t)$ and IMF mode $C_i(t)$, one can design a $\omega$-conditional statistics for all IMF modes as, i.e.,
\begin{equation}
\mathcal{L}_q(\omega)=\langle C^q_i(t)\vert \omega_{i}\rq{}(t)=\omega\rangle_{i,t}\sim \omega^{-\zeta^{H}(q)},
\end{equation}
in which $\zeta^H(q)$ is a scaling exponent comparable with $\zeta^S(q)$. 
It is found that the generalized HHT can suppress the effect of the energetic structure to retrieve the real scaling or singularity spectrum~\cite{Huang2011PRE,Huang2013PRE,Tan2014PoF}. For more detail of this Hilbert-based method, see Refs. \cite{Schmitt2016Book,Huang1998EMD,Huang1999EMD}.

\subsubsection{Multi-level segment analysis}
The multi-level segment analysis (MSA) focuses on the flow structure at different scale levels, which are related to the extremal points of a specified field quantity. Local extrema are determined by both the turbulent random motion and the laminar diffusion. Considering the time series $\tilde{\theta}(t)$, its extrema are conditionally valid. For instance, if $t_0$ is extremal with respect to scale $s$, i.e.
$\tilde{\theta}(t_0) \leq \tilde{\theta}(t), \forall t\in (t_0-s,t_0+s)$ (minimum), or $\tilde{\theta}(t_0)\ge \tilde{\theta}(t), \forall t\in (t_0-s,t_0+s)$ (maximum), it may not be extremal at a larger scale $s_{1}>s$.
For a prescribed $s$, denote the corresponding extremal point set as $t_{s,i}$, $i=1,2,...$. A segment is defined as the part of $\tilde{\theta}(t)$ between two adjacent extremal points. The characteristic parameters to describe the structure skeleton are the function difference $\tilde{\theta}(t_{s,i})-\tilde{\theta}(t_{s,i-1})$ and the time scale $t_{s,i}-t_{s,i-1}$. Scanning over different $s$ to collect all the segment characteristics describes the statistical properties of $\tilde{\theta}(t)$. In this context the structure function (for the $q$th order case) can be defined as
\begin{equation}
\mathcal{D}_q(\tau)=
\langle [\tilde{\theta}(t_i)-\tilde{\theta}(t_{i-1})]^{q}\vert _{t_{i}-t_{i-1}=\tau}\rangle_{s},
\label{newsf}
\end{equation}
where $\langle\cdot\rangle_{s}$ denotes sampling over different $s$. It has been argued that based on the natural topology of the physics process, MSA is effective in resolving multi-scale relations. More technical details of this method can be referred to Ref.~\cite{Wang2015JSTAT}.

\red{Note that both the HHT and MSA methods define  the scale locally to avoid the scale mixing problem~\cite{Huang2011PRE,Huang2010PRE}. For example,  in HHT, the characteristic scale is defined as the distance of two successive extremal points \cite{Huang2014TSC}. While in MSA, scalar is defined as the distance between two consecutive extremal points at specific window sizes. 
Such definition of scale can avoid scale mixing problem: 
the detected scale is derived from the data itself, not arbitrary \textit{a priori} defined. For example, giving a pure sine wave with a fixed frequency, both HHT and MSA identify only one single scale just at the given frequency. While the scale provided by SF analysis is  continuous  from the sampling frequency to the length of the data.} The extreme point plays an important role in both the EMD algorithm and the MAS, which to some extent are related. The multiscaling property \red{seems to be} deeply related with the distribution of these extrema points, which is an interesting topic that beyond of this work \cite{Huang2017PREEPD}.

\section{Results and Discussion}

Here results from different methods are compared to understand the turbulence physics in the present context. In addition, some more general issues will also be tentatively discussed.

\subsection{Intensity of Intermittency}\label{sec:singularity}
\begin{figure}
\centering
\includegraphics[width=1\linewidth,clip]{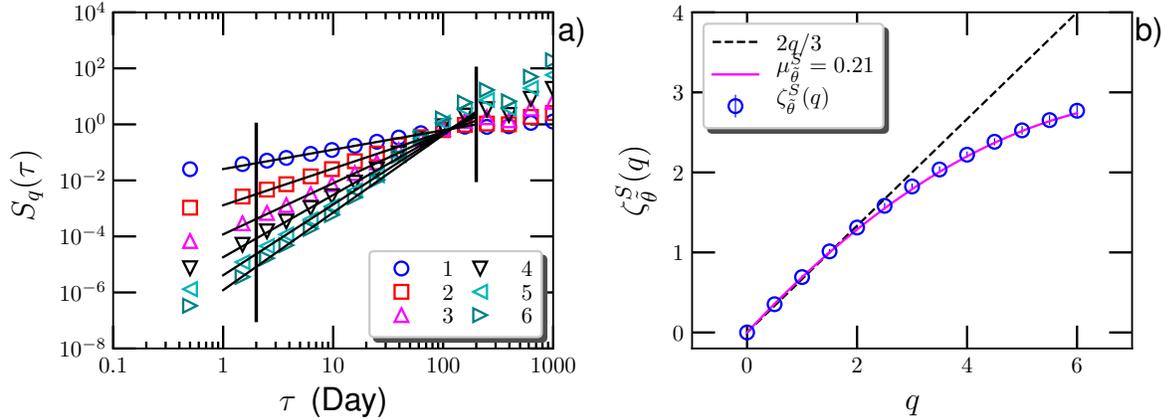}
 \caption{a)  The experimental $q$th-order structure-function $S_q(\tau)$. Power-law behavior is observed in the range $2<\tau<100\,$Day (corresponding to $0.01<f<0.5\,$Day$^{-1}$). b) The measured scaling exponent $\zeta^S_{\tilde{\theta}}(q)$. For comparison, the lognormal formula fit with an intermittency parameter $\mu^S_{\tilde{\theta}}=0.21$ is shown as solid line.}
 \label{fig:qSF}
 \end{figure}

Figure \ref{fig:qSF}\,a) shows the measured high-order statistics of the structure-functions $S_q(\tau)$ from $q=1$ to $q=6$. Power-law behaviour is observed for all $q$ considered here in the range $2<\tau<200\,$Day. The corresponding scaling exponent $\zeta_{\tilde{\theta}}^S(q)$ is retrieved by least square fitting. Figure \ref{fig:qSF}\,b) shows the measured $\zeta_{\tilde{\theta}}^S(q)$ ($\ocircle$), in which the errorbar indicates a $95\%$ fitting confidence interval. For comparison, a linear scaling $\zeta(q)=2q/3$ is presented as a dashed line. Graphically, the convex curve indicates intermittency or multifractality, one of the essential features of turbulence or other turbulence-like dynamical systems. To characterize the intensity of the multifractality or intermittency, we introduce here a lognormal formula, i.e.,
\begin{equation}
\zeta_{\tilde{\theta}}(q)=qH-\frac{\mu_{\tilde{\theta}}}{2}\left(q^2H^2-qH\right),\label{eq:lognormal}
\end{equation}
where $H$ is the Hurst number, and  $\mu_{\tilde{\theta}}$ is the so-called intermittency parameter \cite{Li2014PhysicaA}. Specifically, a larger value $\mu$ has, then more intermittent the field is. We first fix $H=2/3$ (see discussion in Sec.\,\ref{sec:DA}) and then fit $\zeta(q)$ using the above lognormal formula. It yields an intermittency parameter $\mu_{\tilde{\theta}}^S=0.21\pm0.01$. Note that the lognormal formula is first proposed by Kolmogorov in his work in 1962 \cite{Kolmogorov1962} for the Eulerian velocity, i.e.,
\begin{equation}
\zeta_E(q)=\frac{q}{3}-\frac{\mu_E}{2}\left(\frac{q^2}{9}-\frac{q}{3}\right),\label{eq:Klognormal}
\end{equation}
in which $\mu_E$ is the intermittency parameter of the Eulerian velocity. 
\red{The physical hypothesis behind this model is that the energy dissipation field follows the lognormal distribution, which has been reported also valid for the oceanic flow \cite{Pearson2018PRL}.}
A typical experimental value of this parameter for the 3D Eulerian turbulent velocity is $0.2\le \mu_E\le 0.4$ \cite{Frisch1995}. Note that equation \ref{eq:Klognormal} is a special case of equation \ref{eq:lognormal} with $H=1/3$. As shown above, SFs are strongly influenced by the annual cycle. The measured scaling $\zeta_{\tilde{\theta}}^S(q)$ and $\mu_{\tilde{\theta}}^S$ could be biased.

\begin{figure}
\centering
\includegraphics[width=1\linewidth,clip]{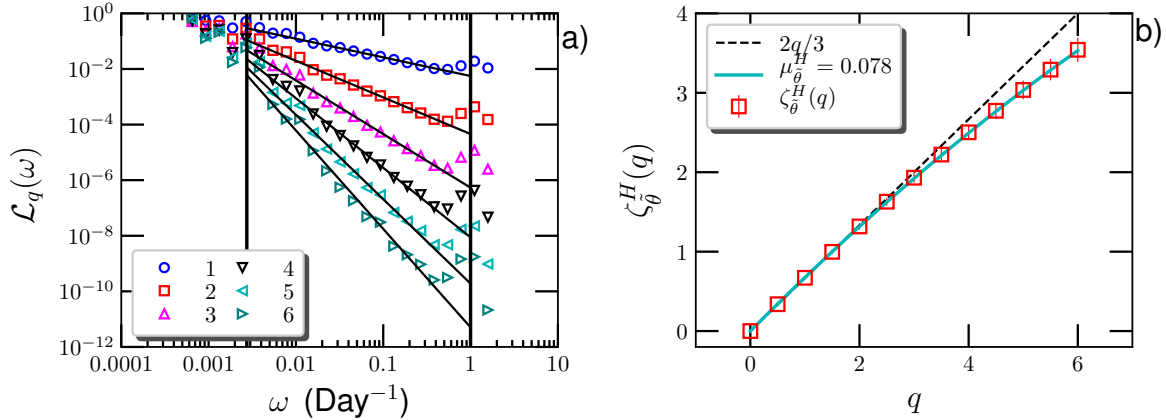}
 \caption{a)  The experimental $q$th-order Hilbert moment $\mathcal{L}_q(\omega)$. Power-law behavior is observed in the range $0.005<\omega<0.5\,$Day$^{-1}$ (corresponding to $2<\omega<200\,$Day). b) The measured scaling exponent $\zeta^H_{\tilde{\theta}}(q)$. For comparison, the lognormal formula fit with an intermittency parameter $\mu^H_{\tilde{\theta}}=0.078$ is shown as solid line.}
 \label{fig:qHilbert}
 \end{figure}

Figure \ref{fig:qHilbert}\,a) shows the experimental  Hilbert marginal spectra $\mathcal{L}_q(\omega)$  from $q=1$ to $6$. A clear power-law behaviour is observed on the same range as the Fourier power spectrum, i.e., $0.005<\omega<0.5\,$Day$^{-1}$. Figure \ref{fig:qHilbert}\,b) shows the fitted scaling exponent $\zeta_{\tilde{\theta}}^H(q)$ ($\square$), in which the errorbar indicates the $95\%$ fitting confidence interval. The corresponding intermittency parameter is estimated as $\mu_{\tilde{\theta}}^H=0.078\pm0.001$ via  the  lognormal formula  equation \ref{eq:lognormal}.  The Hilbert method can isolate the influence of the energetic structures \cite{Huang2013PRE,Tan2014PoF,Huang2009Hydrol,Huang2010PRE}, e.g., the annual cycle in the present data set, and thus provide a better estimation of the scaling exponent $\zeta^H_{\tilde{\theta}}(q)$ and the intermittency parameter $\mu_{\tilde{\theta}}^H$.

\begin{figure}
\centering
\includegraphics[width=1\linewidth,clip]{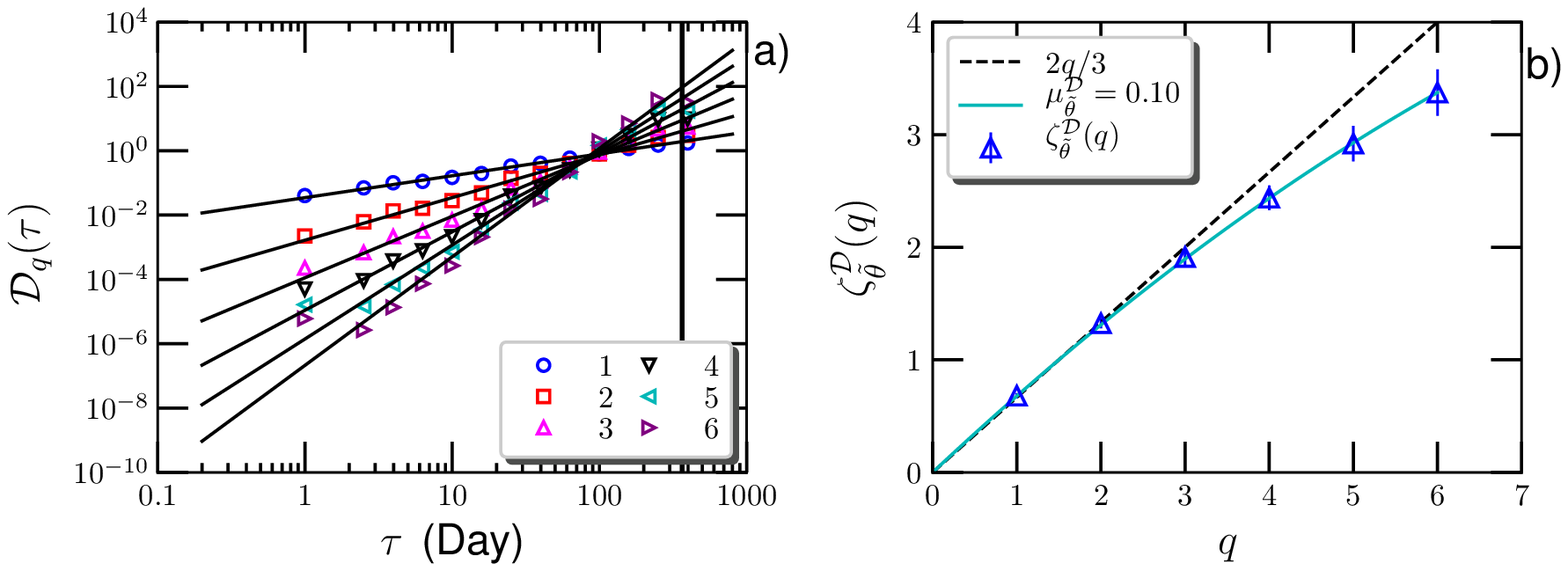}
 \caption{a)  The experimental $q$th-order $\mathcal{D}_q(\ell)$. Power-law behavior is observed in the range $2<\ell<100\,$Day (corresponding to $0.01<\ell<0.5\,$Day$^{-1}$). b) The measured scaling exponent $\zeta^\mathcal{D}_{\tilde{\theta}}(q)$. For comparison, the lognormal formula fit with an intermittency parameter $\mu^{\mathcal{D}}_{\tilde{\theta}}=0.10$ is shown as solid line.}
 \label{fig:qMSA}
 \end{figure}

Figure \ref{fig:qMSA}\,a) shows the measured $\mathcal{D}(\tau)$ for $q=1$ to $6$ using MSA. The power-law behaviour is observed in the range $2<\tau<100\,$Day. The corresponding  measured $\zeta_{\tilde{\theta}}^{\mathcal{D}}(q)$  is shown in Figure \ref{fig:qMSA}\,b) as $\triangle$. It is found that the experiment intermittency parameter $\mu_{\tilde{\theta}}^{\mathcal{D}}=0.10\pm0.01$, which is close to the HHT result.

\begin{figure}
\centering
\includegraphics[width=1\linewidth]{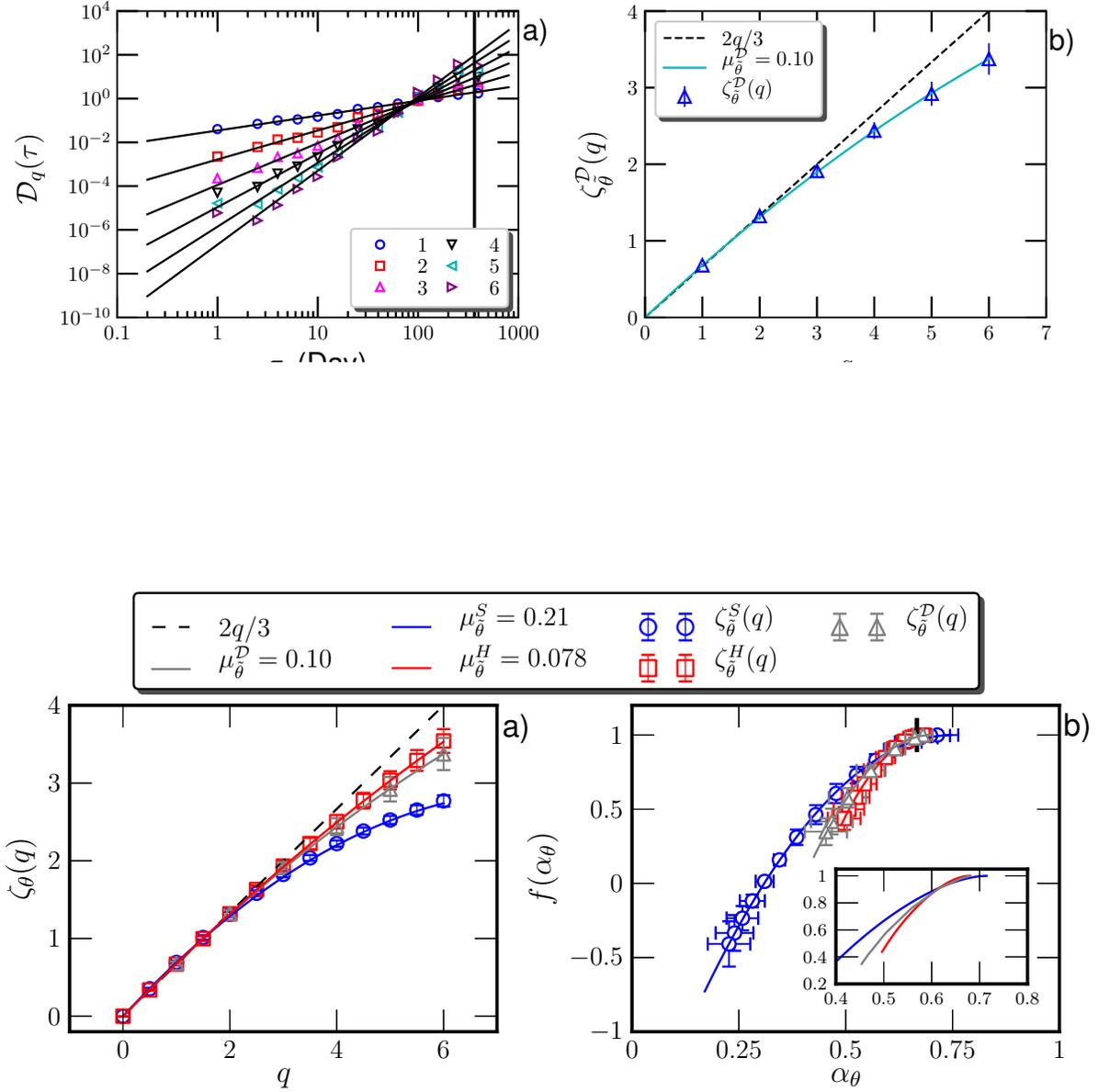}
\caption{a) Comparison of the scaling exponents $\zeta_{\tilde{\theta}}(q)$ determined by three methods ($\ocircle$ from SF, $\square$ from HHT and $\triangle$ from MSA). Solid lines present the corresponding lognormal fits. b) The singularity spectrum $f(\alpha_{\tilde{\theta}})$ for the three different results, with the $95\%$ confidence limit errorbars. The inset shows the enlarged $0.4<\alpha_{\tilde{\theta}}<0.8$ range.}
\label{fig:singularity}
\end{figure}

Figure \ref{fig:singularity}\,a) collectively shows the scaling exponent $\zeta_{\tilde{\theta}}(q)$ from these three methods. Overall HHT and MSA provide almost the same scaling dependence on $q$ and the intermittency parameters are comparable ($\mu_{\tilde{\theta}}\simeq0.10$). For SFs, when $q\le 2$ there is almost no difference, while when $q>2$ the SF scaling exponents show large deviation: i.e. the SF curve bends down, indicating stronger intermittency, which is also indicated by the intermittency parameter.

For 3D homogeneous and isotropic turbulence, $\mu$ for the Eulerian velocity is found to be around $0.2\le \mu_E\le 0.40$. A widely accepted value is $\mu_E\simeq 0.20 $ \cite[see p. 165]{Frisch1995}, which seems to be consistent with the SF estimation $\mu^S_{\overline{\theta}}=0.21$. However, as it has been discussed, the measured SFs, especially for the high-order ones, are strongly influenced by the energetic large-scale structures with the time scale $t\ge T_Y$ (see the figure \ref{fig:2SF}\,b), which is now recognized as the infrared effect \cite{Huang2013PRE}. Therefore, the scaling exponent $\zeta_{\tilde{\theta}}^S(q)$ and the intermittency parameter $\mu_{\tilde{\theta}}^{S}$ determined by the SF approach, are inevitably biased.

To further quantify such difference, we calculate the singularity spectrum $f(\alpha_{\tilde{\theta}})$ to describe the process multifractality via the Legendre transform \cite{Frisch1995} as
\begin{equation}
f(\alpha_{\tilde{\theta}})=\min_{q}(\alpha_{\tilde{\theta}} q-\zeta_{\tilde{\theta}}(q)+1),\quad \alpha_{\tilde{\theta}}=\frac{d\zeta_{\tilde{\theta}}(q)}{dq}.
\end{equation}
Figure~\ref{fig:singularity}\,b) shows the measured $f(\alpha_{\tilde{\theta}})$ with an enlargement inset in the range $0.4<\alpha_{\tilde{\theta}}<0.8$. Considering the $95\%$ confidence limit errorbar, the HHT and MSA curves are comparable, while the SF curve is more broad, implying stronger multifractality.
\red{The lack of the right side of the singularity spectrum $f(\alpha_{\tilde{\theta}})$ is  due to the fact that the negative moments for all these three methods are not feasibly calculated. In principle, the moments $q$ can be in the range $-1\le q \le q_c$, where $q_c$ is the highest order that can be reached by the dataset. However, in practice it requires a large sample size for safe convergence of $q<0$ cases. }

It is worthy noting that to use the appropriate approaches is crucial to extract the \lq\lq{}true\rq\rq{} turbulence physics. Classical SF analysis mixes the information from different scales; while both HHT and MSA are effective in scaling separation. For example, in passive scalar turbulence, SF is strongly influenced by a large-scale ramp-cliff structure \cite{Huang2011PRE}. If the ramp-cliff contamination is confined, one can retrieve the same scaling exponent of the velocity field \cite{Huang2011PRE}, as from the scaling analysis. Another example is the vorticity in 2D turbulence. SF in the forward enstrophy cascade is dominated by the energetic structures in high intensity vorticity events with a spatial size of the injection scale~\cite{Tan2014PoF}. A similar cumulative function analysis, see equation\,(\ref{eq:cumulative})~\cite{Huang2011PRE,Huang2013PRE,Tan2014PoF}, suggests that the scaling from SF analysis can be dominated by such energetic structures in various turbulent systems; thus the results are biased. As pointed out in Ref. \cite{Wang2006JFM} for the general correlation property, in the vicinity of extremal points the two-point correlation and the corresponding scaling behavior are fundamentally different from other regions by nature. Equally averaging with respect to all the spatial points, as in SF, mixes different scaling properties, making the clear scaling range much reduced. Except that the Reynolds number is large enough, scaling laws will be largely contaminated. For instance, it is usually believed that the Lagrangian statistics are strongly Reynolds number dependent, which can be ascribed to the vortex trapping process (i.e. the ultraviolet effect)~\cite{Huang2013PRE}. By conditionally separating different scaling regions, HHT~\cite{Huang2013PRE}, MSA~\cite{Wang2015JSTAT} and the relevant Lagrangian trajectory segment method~\cite{Wang2014PoF} have successfully verified the predicted scaling relation.

\subsection{A Lagrangian Enstrophy-Like Forward Cascade}\label{sec:DA}
As aforementioned, in the complex ocean turbulence system SST can share both 2D and 3D turbulence features. Some tentative analysis is enlightening to understand this problem more quantitatively.

\begin{figure}
\centering
\includegraphics[width=1\linewidth,clip]{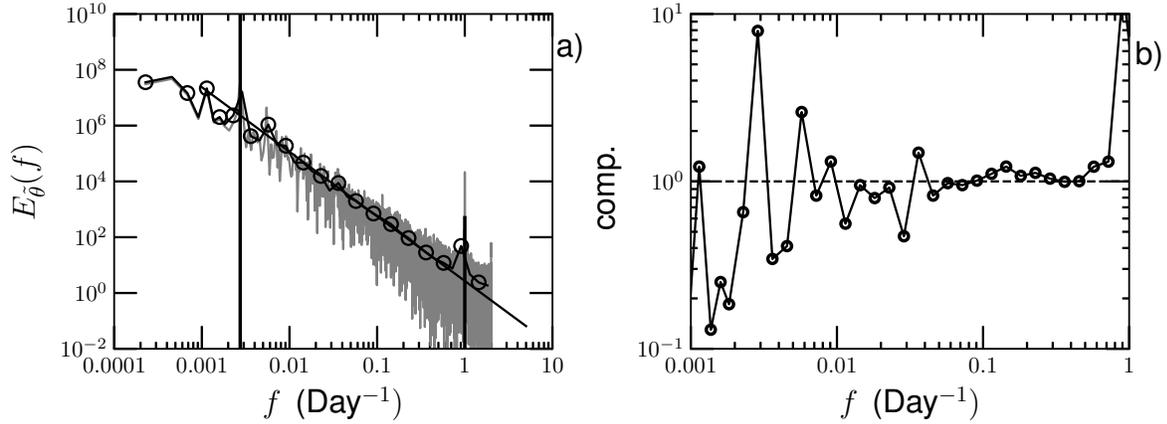}
\caption{a)   The Fourier power spectrum $E_{\tilde{\theta}}(f)$ of $\tilde{\theta}$, in which the solid vertical lines indicate the daily and annual cycles. The symbol is a Fourier power spectrum with 10 bins average each decade in a logarithm scale.
The power-law can be observed in an almost two-decade range $0.005<f<0.5\,$day$^{-1}$ (corresponding to $2<\tau<200\,$day) with a scaling exponent $\beta_{\tilde{\theta}}=2.32\pm0.13\sim7/3$, which is indicated by a solid line. b) The  compensated curve using a fitted parameter to emphasize the observed scaling behavior.  }
\label{fig:SSTspectrum}
\end{figure}

Figure \ref{fig:SSTspectrum}\,a) plots $E_{\tilde{\theta}}(f)$, the Fourier power spectrum of $\tilde{\theta}$ (gray solid line), where the symbol $\ocircle$ is a bin average curve with 10 bins each decade in the logarithm scale to emphasize the power-law behavior. Power-law with a scaling exponent
$\beta_{\tilde{\theta}}=2.32\pm0.13$ can be observed in a two-decade scale range $0.005<f<0.5\,$day$^{-1}$, corresponding to a time scale of $2<\tau<200\,$day. A similar  scaling exponent $\beta=2.44$ has been reported for the Eulerian velocity by \cite{Abraham2002Chaos}. Two vertical solid lines show a daily cycle (the bump at $f\simeq 1\,\si{day^{-1}}$) and annual cycle (the bump around $f\simeq 0.0027\,\si{day^{-1}}$), respectively. To emphasize the observed scaling behavior, Figure\,\ref{fig:SSTspectrum}\,b) shows the compensated curve in a semi-log plot. A visible plateau confirms the existence of the power-law behavior. Note that the scaling exponent $\beta_{\tilde{\theta}}=2.32$ is close to the value $7/3$, which could be obtained via the following tentative dimensional argument.

First simplify the temperature as a passive scalar. Generally the energy cascade can be both forward and inverse. Analogous to K41, it is reasonable to assume that the Fourier spectrum $E_{\tilde{\theta}}(f)$ in the Lagrangian framework is determined by the thermal dissipation $\epsilon_{\tilde{\theta}}$, energy dissipation rate $\epsilon_v$, enstrophy (i.e. the square of vorticity $\Omega$) dissipation $\epsilon_{\Omega}$ and the frequency $f$. Then dimensional analysis yields a temperature spectrum of the following form
\begin{equation}
E_{\theta}(f)\simeq C_{\theta}\epsilon_{\theta}\epsilon_{\Omega}^{1/9}\epsilon_v^{0} f^{-7/3},\label{eq:dimension}
\end{equation}
in which $C_{\theta}$ is a Kolmogorov-like constant. Such $7/3$ scaling agrees with the result shown in the Figure\,\ref{fig:SSTspectrum}\,a). Since here the enstrophy dissipation $\epsilon_{\Omega}$, instead of the energy dissipation rate $\epsilon_v$,  is a relevant parameter to determine the energy spectrum, this result can tentatively be considered as an evidence of an enstrophy-like forward cascade of SST. To our best knowledge, the enstrophy-like cascade is the
signature of the 2D or 2D-like turbulence~\cite{Boffetta2012ARFM,Bouchet2012PhysRep}. It can further be argued that a forward cascade mechanism exists in the scaling range since the main injection scale is around $1\,$year.

A generalization of the above dimensional argument for $q$th-order statistics predicts a nonintermittent scaling behavior, i.e.,
\begin{equation}
M_q(\tau)\sim \epsilon_{\theta}^{q/2}\epsilon_{\Omega}^{q/18}\epsilon_v^0\tau^{\zeta_{\tilde{\theta}}(q)},\,\zeta_{\tilde{\theta}}(q)=2q/3.\label{eq:dimensionQ}
\end{equation}
$M_q(\tau)$ can represent  one of $S_q(\tau)$, $\mathcal{L}_q(\omega)$ or $\mathcal{D}_q(\tau)$. This is the reason why we choose $H=2/3$ in the lognormal formula, see equation\,\ref{eq:lognormal}.

It has been debated for a long time whether the forward enstrophy cascade is intermittent or not \cite{Paret1999PRL,Nam2000PRL}. One difficulty for the final conclusion is that SF fails to detect the scaling behavior of the forward enstrophy cascade. More recently, Tan et al.~\cite{Tan2014PoF} applied the same Hilbert-based analysis to the vorticity field of the 2D turbulence. The measured scaling exponent and singularity spectrum confirm the
multifractality of the forward enstrophy cascade. In equation\,\ref{eq:dimensionQ} the $q/18$ power of $\epsilon_{\Omega}$ can be a possible reason why intermittency of the SST field is weak.

The bump in Figure\,\ref{fig:SSTspectrum}\,a) indicates another energy injection scale around $1\,$day. The temperature fluctuation is then transferred from $1\,$day to smaller scales, e.g. few minutes or seconds, through a 3D forward energy cascade~\cite{Frisch1995}. Meanwhile, such energy can also be transferred to large scales via the inverse cascade mechanism to organize large-scale structures~\cite{Xia2011NatPhys}. Therefore, the cascade direction between $1\,$day and $1\,$year is then a result of the competition between the forward enstrophy-like cascade and the
inverse energy-like cascade~\cite{Celani2011PRL}. From the present $7/3$ scaling result, it seems that below the annual cycle the forward enstrophy-like cascade dominates. Above the annual cycle, the inverse cascade may lead to a system-size structure \cite{Xia2011NatPhys}, such as the Pacific decadal oscillation.

\section{Conclusions}
\red{To investigate turbulence and other generally complex systems, more advanced analysis methods need to be developed to extract the important process physics. For ocean turbulence, under the joint action of fluid dynamics, geophysics and the external atmospheric influences, the sea surface temperature (SST) problem is extremely challenging. In the present work, we focus on the scaling and intermittency properties of the averaged SST as a time series from the floater measurement results, which inherits the important turbulent features.} We interpret the averaged SST as the Lagrangian or partially filtered Lagrangian temperature based on the fractal dimension feature of the spatial distribution of drifters. Mainly we introduce the different methods, including the Hilbert-Huang transform (HHT) and multi-level segment analysis (MSA), to investigate the scaling behavior of SST. It has been found that the conventional structure function approach inevitably mixes the scaling relations at different scales. Thus the results are biased, showing stronger intermittency. In contrast, both HHT and MSA by nature are effective in scale separating. Therefore scaling mixing can be much reduced to reveal a weaker intermittency intensity.

Ocean turbulence is more complex than the canonical cases due to different involved factors, e.g., inhomogeneity, anisotropic, complex boundary, complex external forcing, stratification, waves, (weak) compressibility, etc. Dimensional argument based on the energy spectrum  of SST suggests a two-dimensional-like Lagrangian forward cascade in which the enstrophy dissipation $\epsilon_{\Omega}$ is involved. The high-order generalization is also confirmed by the data with intermittency correction. However, due to the system complexity and the limited data, it is hard to conclude exactly which theory, such as Kraichnan two-dimensional turbulence theory, geophysical turbulence, etc.,  should be recommended to explain the obtained results here. 
Some key issues need to be further investigated, such as the cascade direction, passive or active, etc., by taking into account more involved important factors. The present work may inspire some new theoretical and more comprehensive considerations to understand ocean turbulence.

\section*{Acknowledgments}
This work is sponsored by the National Natural Science Foundation of China (under Grant Nos. 11332006, 11732010 and 91441116), and  partially by the Sino-French (NSFC-CNRS) joint research project (No. 11611130099, NSFC China, and PRC 2016-2018 LATUMAR ``Turbulence lagrangienne: \'etudes num\'eriques et applications environnementales marines",  CNRS, France).  Y.H. is also supported by the Fundamental Research Funds for the Central Universities (Grant No. 20720180120 and 20720180123), and MEL Internal Research Fund (Grant No. MELRI1802). Y.H. thanks  Dr. G. Rilling and Prof. P. Flandrin from laboratoire de Physique, CNRS \& ENS Lyon (France) for sharing their Empirical Mode Decomposition (EMD) \textsc{Matlab} codes, which is available at: {http://perso.ens-lyon.fr/patrick.flandrin/emd.html}. A source package to realize the Hilbert spectral analysis and multi-level segment analysis is available at: {https://github.com/lanlankai}.

\pagebreak

\bibliographystyle{iopart-num}
\providecommand{\newblock}{}

\end{document}